\documentclass{article}
\usepackage{amsfonts}
\usepackage{amssymb}
\usepackage{amsmath}
\usepackage{graphicx}

\begin{document}

\title{Group actions as stroboscopic maps of ordinary differential equations}
\author{Andrzej Okninski \\
Politechnika Swietokrzyska, Physics Division,\\
Al. 1000-lecia PP 7, 25-314 Kielce, Poland}
\maketitle

\begin{abstract}
Discrete-time dynamical systems can be derived from group actions. In the
present work possibility of application of this method to systems of
ordinary differential equations is studied. Invertible group actions are
considered as possible candidates for stroboscopic maps of ordinary
differential equations. It is shown that flow of the Bloch equation is a
unique suspension of an invertible map on the $SU\left( 2\right) $ group.
\end{abstract}

\section{Introduction}

Discrete-time dynamical systems can be formulated in terms of group actions
to exploit the group structure and get a better understanding of the
corresponding dynamics. This approach was used to study non-invertible
discrete-time dynamics on $E\left( 2\right) $ \cite{Okninski1992} and $%
SU\left( 2\right) $ \cite{Okninski1992,Okninski1994,Ahmed1994,Okninski1996a}%
. The Shimizu-Leutbecher map \cite{Shimizu1963,Leutbecher1967}, a tool to
study group structure, was solved for an arbitrary Lie group $\mathcal{G}$ 
\cite{Okninski1992}. Since the Shimizu-Leutbecher sequence generates the
logistic map for $\mathcal{G}=SU\left( 2\right) $ \cite%
{Okninski1992,Ahmed1994} a general solution of the logistic map was thus
obtained. For $\mathcal{G}=E\left( 2\right) $ this approach led to a better
understanding of the Harter-Heighway fractal curve \cite{Okninski1992}. A
two-dimensional generalization of the logistic map was also introduced as a
map on $SU\left( 2\right) $\ and investigated \cite{Okninski1996a}.

On the other hand, structure of Kleinian groups is naturally studied in the
setting of discrete-time dynamical systems, revealing in this way
connections with fractals \cite{Brooks1981,Gehring1989,Mumford2002}. There
were several attempts to study dynamics on Lie groups within continuous-time
rather than discrete-time approach. Continuous-time dynamics on $SU\left(
2\right) $ and $SU\left( 2\right) \times SU\left( 2\right) $ was defined in
the setting of ordinary differential equations (ODEs) \cite%
{Kowalski1996,Kowalski1998} while continuous iteration of maps was defined
to find a correspondence between mappings and continuous-time evolution \cite%
{Aldrovandi1998,Gralewicz2000}.

The question now arises whether a general connection between discrete-time
group actions and continuous-time dynamical systems can be established.

We investigate a possibility of relating group actions with stroboscopic
maps of ODEs. Let us consider a continuous-time dynamical system given by a
set of ODEs 
\begin{equation}
\frac{d\mathbf{x}}{dt}=\mathbf{f}\left( \mathbf{x}\left( t\right) ,t\right) ,
\label{ODE}
\end{equation}
where $\mathbf{x}=\left[ x_{1},\ x_{2},\ \ldots \ ,\ x_{n}\right] ,$ $%
\mathbf{f}=\left[ f_{1},\ f_{2},\ \ldots \ ,\ f_{n}\right] $. Let $\mathbf{x}
\left( t\right) $ be a solution to Eq.(\ref{ODE}). Then the invertible map 
\begin{equation}
S_{T}:\quad \mathbf{x}\left( t\right) \longmapsto \mathbf{x}\left(
t+T\right) ,  \label{strobe}
\end{equation}
is a stroboscopic map of strobe time $T$. Stroboscopic maps with several
strobe times are standard tools to solve differential equations to mention
only the Runge-Kutta methods. On the other hand, a discrete map can be also
used to generate a continuous flow. Such a flow, non unique in general, is
called a suspension of the map \cite{Tufillaro1992,deVries1993}.

In the present work we shall consider a class of invertible maps on a group
and then try to find a flow, preferably unique, such that its trajectories
connect continuously the iterates of the map.

The paper is organized as follows. In the next Section maps on a Lie group $%
\mathcal{G}$ are defined and method to deduce evolution of group parameters
is described. Principal results are described in Sections 3 and 4. In
Section 3 a class of invertible maps is considered. These maps are solved in
Subsection 3.1. In Subsection 3.2 a simplified map is defined, solved and
parameterized on the $SU\left( 2\right) $ group. In Subsection 3.3 symmetry
and restrictions of the dynamics are determined. Results of the Subsection
3.2 are used in Section 4 to derive ODE for which the simplified map is a
stroboscopic map. The map samples the flow of the ODE exactly and
arbitrarily densely and it follows that the construction is unique. This ODE
is the Bloch equation \cite{Allen1975}, also known as the Landau-Lifshitz
equation. These results generalize an earlier findings \cite{Okninski1996b}.
In Section 5 computations for the simplified map on the $SU\left( 2\right) $
group are presented to elucidate dynamics of the Bloch equation. In the last
Section the obtained results are summarized and perspectives of further
research are outlined.

\section{Group dynamical systems}

Let us consider a dynamical system defined by the following map 
\begin{equation}
G_{N+1}=\varphi \left( G_{N},\ \ldots \right) ,  \label{groupdynamics}
\end{equation}%
where $G_{N}\in \mathcal{G}$. Let $\mathcal{G}$ be a simple Lie group and $%
\mathfrak{g}$ its Lie algebra. Then $G_{N}\in \mathcal{G}$ can be written in
exponential form%
\begin{equation}
G_{N}=\exp \left( X_{N}\right) ,  \label{expform}
\end{equation}%
where $X_{N}\in \mathfrak{g}$ \cite{Barut1977}. Any infinitesimal operator $%
X_{N}$ of a $n$-parameter Lie group $\mathcal{G}$ is a linear combination of 
$n$ generators $I^{k}$%
\begin{equation}
X_{N}=\sum_{k=1}^{n}I^{k}c_{N}^{k},\quad I^{k}\in \mathfrak{g}\mathbf{,}
\label{groupelement}
\end{equation}%
where real parameters $c_{N}^{1},...,c_{N}^{n}$ are local coordinates of the
Lie group element $X_{N}$.

Substituting Eqs.(\ref{expform}), (\ref{groupelement}) into Eq.(\ref%
{groupdynamics}) and using completeness of the basis consisting of the
generators $I^{k}$\ and the unit matrix $\mathbf{1}$\ we get a discrete-time
dynamical system in parameter space \cite{Okninski1992,Gajdek1995}%
\begin{equation}
c_{N+1}^{j}=F^{j}\left( c_{N}^{1},\ \ldots ,\ c_{N}^{n}\right) ,\quad j=1,\
\ldots ,\ n,  \label{parameterdynamics}
\end{equation}%
where $F^{j}$ are continuous functions \cite{Pontryagin1966}.

\section{Discrete-time dynamics on a group}

Let us consider an invertible discrete-time dynamical system on a Lie group $%
\mathcal{G}$ 
\begin{equation}
R_{N+1}=Q_{N}R_{N}Q_{N-1}R_{N-1}Q_{N-1}^{-1}R_{N}^{-1}Q_{N}^{-1},\qquad
N=1,2,\ldots ,  \label{QRQ}
\end{equation}%
i.e. $\varphi =Q_{N}R_{N}Q_{N-1}R_{N-1}Q_{N-1}^{-1}R_{N}^{-1}Q_{N}^{-1}$ in
Eq.(\ref{groupdynamics}), where we assume knowledge of all group elements $%
Q_{M}$ needed for the computations. Let us note here that apparently simpler
non-invertible Shimizu-Leutbecher map $R_{N+1}=R_{N}QR_{N}^{-1}$ has a
complicated solution \cite{Okninski1992}. It turns out, however, that a
relatively simple solution to (\ref{QRQ}) can be constructed upon
introducing new variables $S_{N}$ defined below.

\subsection{Exact solution}

We note that Eq.(\ref{QRQ}) can be reduced to two simpler equations. Indeed,
introducing new quantity 
\begin{equation}
S_{N}\overset{df}{=}R_{N}Q_{N-1}R_{N-1}Q_{N-2},  \label{defS}
\end{equation}%
we can rewrite Eq.(\ref{QRQ}) as 
\begin{subequations}
\label{RS}
\begin{align}
R_{N+1}& =S_{N+1}R_{N-1}S_{N+1}^{-1},\quad N=1,2,\ldots \ ,  \label{R} \\
S_{N+1}& =Q_{N}S_{N}Q_{N-2}^{-1},\hspace{1.05cm}N=1,2,\ldots \ .  \label{S}
\end{align}

To run dynamics defined by (\ref{QRQ}) or, alternatively, by (\ref{RS}), we
have to impose initial conditions for Eq.(\ref{QRQ}), $R_{0},\ R_{1}$, from
which initial condition for Eq.(\ref{S}), $S_{2}=R_{2}Q_{1}R_{1}Q_{0}$, can
be also computed. Equations (\ref{RS}) are easily solved 
\end{subequations}
\begin{subequations}
\label{SOL}
\begin{align}
R_{2K}& =S_{2K}S_{2K-1}\ldots S_{2}R_{0}S_{2}^{-1}\ldots
S_{2K-1}^{-1}S_{2K}^{-1},  \label{solR} \\
S_{N}& =Q_{N-1}\ldots Q_{1}S_{1}Q_{-1}^{-1}\ldots Q_{N-3}^{-1},  \label{solS}
\end{align}%
where $K=1,2,\ldots $\ , $N=2,3,\ldots $ and similar equations can be
written for $R_{2K+1}$.

\subsection{Discrete-time dynamics on the sphere}

In the case $\mathcal{G}=SU(2)$ the Hamilton's parameterization can be used.
Let unit vectors $\mathbf{r}_{N}=\left( r_{N}^{1},\ r_{N}^{2},\
r_{N}^{3}\right) $,$\ \mathbf{q}_{N}=\left( q_{N}^{1},\ q_{N}^{2},\
q_{N}^{3}\right) $ and angles $\chi _{N}$, $\alpha _{N}$ correspond to
rotation axes and rotation angles, respectively. Then the rotation matrices $%
R_{N},\ Q_{N}$ are defined as 
\end{subequations}
\begin{eqnarray}
R_{N} &=&\exp \left( i\tfrac{\chi _{N}}{2}\mathbf{\sigma }\cdot \mathbf{r}%
_{N}\right) ,\qquad \left\vert \mathbf{r}_{N}\right\vert =1,  \label{defR} \\
Q_{N} &=&\exp \left( i\tfrac{\alpha _{N}}{2}\mathbf{\sigma }\cdot \mathbf{q}%
_{N}\right) ,\qquad \left\vert \mathbf{q}_{N}\right\vert =1,  \label{defQ}
\end{eqnarray}%
where $i^{2}=-1$, and $\mathbf{\sigma }=\left[ \sigma ^{1},\ \sigma ^{2},\
\sigma ^{3}\right] $ is the pseudo vector with the Pauli matrices as
components \cite{Misner1973}. With parameterization (\ref{defR}), (\ref{defQ}
) Eq.(\ref{QRQ}) induces dynamics of vectors $\mathbf{r}_{N}$ on unit
sphere. We shall consider a special case $Q_{N}\equiv Q$ 
\begin{equation}
R_{N+1}=QR_{N}QR_{N-1}Q^{-1}R_{N}^{-1}Q^{-1}.  \label{QRQsimpl}
\end{equation}

The solution of (\ref{QRQsimpl}) is immediately obtained from (\ref{SOL}): 
\begin{subequations}
\label{RSsol}
\begin{align}
R_{2K}& =Q^{2K}P^{K}R_{0}P^{-K}Q^{-2K},  \label{2k} \\
\tilde{R}_{2K+1}& =Q^{2K}P^{K}\tilde{R}_{1}P^{K}Q^{-2K},  \label{2k+1}
\end{align}
where $P\overset{df}{=}Q^{-1}S_{1}Q^{-1}=Q^{-1}R_{1}QR_{0},\ \tilde{R}%
_{2K+1} \overset{df}{=}Q^{-1}R_{2K+1}Q,\ K=0,\ 1,\ 2,\ \ldots \ $. It
follows that equations (\ref{2k}), (\ref{2k+1}) generate analogous dynamics.
Matrices $Q,\ P$ are parameterized as 
\end{subequations}
\begin{subequations}
\label{QP}
\begin{align}
Q& =\exp \left( i\tfrac{\alpha }{2}\mathbf{\sigma }\cdot \mathbf{q}\right)
,\qquad \left\vert \mathbf{q}\right\vert =1,  \label{Q} \\
P& =\exp \left( i\tfrac{\beta }{2}\mathbf{\sigma }\cdot \mathbf{p}\right)
,\qquad \left\vert \mathbf{p}\right\vert =1,  \label{P}
\end{align}
where $\alpha $, $\beta $ are the corresponding angles of rotations while $%
\mathbf{p}=\left[ p^{1},p^{2},p^{3}\right] $ and $\mathbf{q}=\left[
q^{1},q^{2},q^{3}\right] $ are unit vectors. We still have to impose initial
condition $R_{0}$ while $R_{1}$ is computed as $R_{1}=QPR_{0}^{-1}Q^{-1}$.

A sufficient condition that points generated according to (\ref{2k}) form a
periodic trajectory (a finite set of points) is that for some integer $K$
the following conditions hold: 
\end{subequations}
\begin{equation}
K\beta =2m\pi ,\ 2K\alpha =2n\pi ,  \label{periodic}
\end{equation}
for some integer $m,\ n;$ in this case the parameter $\beta /\left( 2\alpha
\right) =m/n$ is rational.

Let us note that due to properties of the Pauli matrices the vector $\mathbf{%
\ x}\left( \gamma \right) $ defined by 
\begin{equation}
\mathbf{\sigma }\cdot \mathbf{x}\left( \gamma \right) =S\ \mathbf{\sigma }
\cdot \mathbf{x\ }S^{-1}=\exp \left( i\tfrac{\gamma }{2}\mathbf{\sigma }
\cdot \mathbf{s}\right) \,\mathbf{\sigma }\cdot \mathbf{x}\,\exp \left( -i 
\tfrac{\gamma }{2}\mathbf{\sigma }\cdot \mathbf{s}\right) ,  \label{defx}
\end{equation}
where $\mathbf{x}=\left[ x^{1},x^{2},x^{3}\right] $, $\mathbf{s}=\left[
s^{1},s^{2},s^{3}\right] $ are unit vectors and $\gamma $ is a corresponding
angle of rotation, is easily computed as 
\begin{equation}
\mathbf{x}\left( \gamma \right) =\cos \left( \gamma \right) \mathbf{x}+\sin
\left( \gamma \right) \mathbf{x}\times \mathbf{s}+\left( 1-\cos \left(
\gamma \right) \right) \left( \mathbf{s}\cdot \mathbf{x}\right) \mathbf{s}.
\label{x}
\end{equation}

Using this result it is possible to write the solution (\ref{2k}) in a
closed form. Indeed, equation (\ref{2k}) can be explicitly written as

\begin{equation}
\mathbf{\sigma }\cdot \mathbf{r}_{2K}=\exp \left( iK\alpha \mathbf{\sigma }
\cdot \mathbf{q}\right) \exp \left( iK\tfrac{\beta }{2}\mathbf{\sigma }\cdot 
\mathbf{p}\right) \,\mathbf{\sigma }\cdot \mathbf{r}_{0}\,\exp \left( -iK 
\tfrac{\beta }{2}\mathbf{\sigma }\cdot \mathbf{p}\right) \exp \left(
-iK\alpha \mathbf{\sigma }\cdot \mathbf{q}\right) \mathbf{.}  \label{RSsol1a}
\end{equation}

Applying Eq.(\ref{x}) twice we find the closed form solution of Eq.(\ref%
{QRQsimpl}): 
\begin{subequations}
\label{EXACTa}
\begin{align}
\mathbf{r}\left( K\alpha \right) & =\cos \left( K\alpha \right) \mathbf{t}
\left( K\alpha \right) +\sin \left( K\alpha \right) \mathbf{t}\left( K\alpha
\right) \times \mathbf{q}+\left( 1-\cos \left( K\alpha \right) \right)
\left( \mathbf{q}\cdot \mathbf{t}\left( K\alpha \right) \right) \mathbf{q,}
\label{r1} \\
\mathbf{t}\left( K\alpha \right) & =\cos \left( \lambda K\alpha \right) 
\mathbf{r}_{0}+\sin \left( \lambda K\alpha \right) \mathbf{r}_{0}\times 
\mathbf{p}+\left( 1-\cos \left( \lambda K\alpha \right) \right) \left( 
\mathbf{p}\cdot \mathbf{r}_{0}\right) \mathbf{p},  \label{r2}
\end{align}
where $\mathbf{r}\left( K\alpha \right) \overset{df}{=}\mathbf{r}_{2K}$, $%
\lambda \overset{df}{=}\beta /\left( 2\alpha \right) $ and $K=0,1,2,\ldots $
.

The sequence of vectors $\mathbf{r}_{0},\ \mathbf{r}_{2},\ \mathbf{r}%
_{4},\ldots ,$ generated according to Eq.(\ref{EXACTa}), samples a
continuous curve $\mathcal{C}$. Indeed, it follows from Eq.( \ref{EXACTa})
that for very small $\alpha $ and fixed $\lambda $ vectors $\mathbf{r}\left(
K\alpha \right) $ and $\mathbf{r}\left( \left( K+1\right) \alpha \right) $
are very close on a unit sphere. Therefore for decreasing $\alpha $, $\beta $
and fixed $\lambda =\beta /\left( 2\alpha \right) $ the sequence $\mathbf{r}%
_{0},\ \mathbf{r}_{2},\ \mathbf{r}_{4},\ldots $ approximates the curve $%
\mathcal{C}$ more and more exactly. It is useful to introduce new variable $%
\theta \overset{df}{=}K\alpha $ which can treated as continuous since $%
\alpha $ is arbitrary. Using this we can rewrite Eq.(\ref{RSsol1a}) as: 
\end{subequations}
\begin{equation}
\mathbf{\sigma }\cdot \mathbf{r}\left( \theta \right) =\exp \left( i\theta 
\mathbf{\sigma }\cdot \mathbf{q}\right) \exp \left( i\lambda \theta \mathbf{%
\ \sigma }\cdot \mathbf{p}\right) \,\mathbf{\sigma }\cdot \mathbf{r}%
_{0}\,\exp \left( -i\lambda \theta \mathbf{\sigma }\cdot \mathbf{p}\right)
\exp \left( -i\theta \mathbf{\sigma }\cdot \mathbf{q}\right) ,
\label{RSsol1b}
\end{equation}%
while Eq.(\ref{EXACTa}) leads to explicit formula for the curve $\mathcal{C}$%
: 
\begin{subequations}
\label{EXACTb}
\begin{align}
\mathbf{r}\left( \theta \right) & =\cos \left( \theta \right) \mathbf{t}%
\left( \theta \right) +\sin \left( \theta \right) \mathbf{t}\left( \theta
\right) \times \mathbf{q}+\left( 1-\cos \left( \theta \right) \right) \left( 
\mathbf{q}\cdot \mathbf{t}\left( \theta \right) \right) \mathbf{q,}
\label{r3} \\
\mathbf{t}\left( \theta \right) & =\cos \left( \lambda \theta \right) 
\mathbf{r}_{0}+\sin \left( \lambda \theta \right) \mathbf{r}_{0}\times 
\mathbf{p}+\left( 1-\cos \left( \lambda \theta \right) \right) \left( 
\mathbf{p}\cdot \mathbf{r}_{0}\right) \mathbf{p}.  \label{r4}
\end{align}

\subsection{Symmetry and restrictions of dynamics}

Dynamical system (\ref{QRQsimpl}) has continuous symmetry: 
\end{subequations}
\begin{equation}
R_{N}\rightarrow Q^{\kappa }R_{N}Q^{-\kappa },\qquad \forall \kappa \in 
\mathbb{R}.  \label{sym}
\end{equation}%
This symmetry is equivalent to rotation of $\mathbf{r}_{N}$ around $\mathbf{q%
}$ about an arbitrary angle $\kappa \alpha $. It can be thus expected that
dynamics of the quantity $\mathbf{r}_{N}\cdot \mathbf{q}$ should decouple
from other degrees of freedom in (\ref{QRQsimpl}) \cite%
{Okninski1994,Okninski1996a}. Indeed, it follows from (\ref{r1}) that 
\begin{equation}
\mathbf{r}\left( \theta \right) \cdot \mathbf{q}=\mathbf{t}\left( \theta
\right) \cdot \mathbf{q.}  \label{decoupling}
\end{equation}

Since $\left\vert \mathbf{t}\left( \theta \right) \right\vert =\left\vert 
\mathbf{q}\right\vert =1$ it follows from the Schwartz inequality that $%
-1\leq \mathbf{t}\left( \theta \right) \cdot \mathbf{q}\leq 1$. Now, for
given $\mathbf{p}$, $\mathbf{q}$ and $\mathbf{r}_{0}$ we have 
\begin{equation}
A_{1}\leq \mathbf{t}\left( \theta \right) \cdot \mathbf{q}\leq A_{2},\qquad
-1\leq A_{1},\quad A_{2}\leq 1.  \label{bounds1}
\end{equation}
The constants $A_{1,2}$ depending on the parameters $\mathbf{p}$, $\mathbf{q}
$ and the initial condition $\mathbf{r}_{0}$ can be computed from (\ref{r2})
by elementary means 
\begin{subequations}
\label{BOUNDS}
\begin{equation}
A_{1,2}=c\mp \sqrt{b^{2}+\left( a-c\right) ^{2}},  \label{bounds2}
\end{equation}
where 
\begin{equation}
a=\mathbf{r}_{0}\cdot \mathbf{q,\quad }b=\left( \mathbf{r}_{0}\times \mathbf{%
\ p}\right) \cdot \mathbf{q,\quad }c=\left( \mathbf{p}\cdot \mathbf{r}
_{0}\right) \left( \mathbf{p}\cdot \mathbf{q}\right) .  \label{bounds3}
\end{equation}

It thus follows that the motion on the sphere $\left\vert \mathbf{r}\left(
\theta \right) \right\vert =1$ is bounded by two parallels: $A_{1}\leq 
\mathbf{r}\left( \theta \right) \cdot \mathbf{q}\leq A_{2}$.

\section{Differential equation}

The map (\ref{QRQsimpl}), as follows from the solution shown in Eq.(\ref%
{EXACTa}), is the stroboscopic map with strobe time $\alpha $ of a
differential equation which will be deduced from the solution (\ref{RSsol1b}
) where $\theta $ is a continuous variable (it can be derived from the form (%
\ref{EXACTb}) as well). Differentiating Eq.(\ref{RSsol1b}) with respect to $%
\theta $ and using (\ref{RSsol}) we get (see Exercise $41.3$ in
Ref.\thinspace \cite{Misner1973} for similar computations) 
\end{subequations}
\begin{equation}
\dfrac{d\,\mathbf{\sigma }\cdot \mathbf{r}\left( \theta \right) }{d\theta }=i%
\left[ \mathbf{\sigma }\cdot \mathbf{u}\left( \theta \right) ,\ \mathbf{\
\sigma }\cdot \mathbf{r}\left( \theta \right) \right] ,  \label{RSODE}
\end{equation}%
where $\left[ A,\ B\right] \overset{df}{=}AB-BA$ and 
\begin{subequations}
\label{DEFS}
\begin{align}
\mathbf{u}\left( \theta \right) & =\mathbf{q}+\lambda \mathbf{p}\left(
\theta \right) ,\quad \lambda =\tfrac{\beta }{2\alpha },  \label{defu} \\
\mathbf{\sigma }\cdot \mathbf{p}\left( \theta \right) & =\exp \left( i\theta 
\mathbf{\sigma }\cdot \mathbf{q}\right) \,\mathbf{\sigma }\cdot \mathbf{p}%
\,\exp \left( -i\theta \mathbf{\sigma }\cdot \mathbf{q}\right) .
\label{defrtheta}
\end{align}

It follows that the sequence $\mathbf{r}_{0},\mathbf{r}_{2},\mathbf{r}%
_{4},\ldots $ , generated from Eq.(\ref{EXACTa}), samples the flow of Eq.(%
\ref{RSODE}) exactly. As it was remarked in Subsection 3.2 the vectors $%
\mathbf{r}_{2K}$ and $\mathbf{r}_{2\left( K+1\right) }$ are very close on a
unit sphere for very small $\alpha $ and fixed $\lambda $, cf. Eq.(\ref%
{EXACTa}). Therefore for decreasing $\alpha $, $\beta $ and fixed $\lambda
=\beta /\left( 2\alpha \right) $ the sequence of points on the unit sphere
given by $\mathbf{r}_{0},\ \mathbf{r}_{2},\ \mathbf{r}_{4},\ldots $ can be
arbitrarily dense.

Equations (\ref{RSODE}), (\ref{DEFS}) can be written in explicit form. Using
Eqs.(\ref{defx}), (\ref{x}) we get 
\end{subequations}
\begin{equation}
\mathbf{p}\left( \theta \right) =\cos \left( 2\theta \right) \mathbf{p}+\sin
\left( 2\theta \right) \mathbf{p}\times \mathbf{q}+\left( 1-\cos \left(
2\theta \right) \right) \left( \mathbf{q}\cdot \mathbf{p}\right) \mathbf{q},
\label{defp}
\end{equation}%
and hence%
\begin{equation}
\dfrac{d\,\mathbf{r}\left( \theta \right) }{d\theta }=2\mathbf{r}\left(
\theta \right) \times \mathbf{u}\left( \theta \right) ,  \label{r}
\end{equation}%
with $\mathbf{u}\left( \theta \right) $ given by Eqs.(\ref{defu}), (\ref%
{defp}). Obviously the length of the vector $\mathbf{r}$ is a conserved
quantity, and we shall put $\left\vert \mathbf{r}\left( \theta \right)
\right\vert =1$. The angle $\theta $ can be treated as increasing with
angular velocity $\omega =const$, $\dfrac{d\theta }{dt}=\omega \ $\cite%
{Misner1973}, and time variable can be introduced to obtain finally

\begin{subequations}
\label{RUT}
\begin{align}
\dfrac{d\,\mathbf{r}\left( t\right) }{dt}& =2\mathbf{r}\left( t\right)
\times \mathbf{u}\left( t\right) ,  \label{rt} \\
\mathbf{u}\left( t\right) & =\omega \left\{ \left[ 1+\lambda -\lambda \cos
\left( 2\omega t\right) \left( \mathbf{q}\cdot \mathbf{p}\right) \right] 
\mathbf{q}+\lambda \cos \left( 2\omega t\right) \mathbf{p}+\lambda \sin
\left( 2\omega t\right) \mathbf{p}\times \mathbf{q}\right\} .  \label{ut}
\end{align}

Let us note that Eq.(\ref{RUT}) is the Bloch equation \cite{Allen1975}.
Equation (\ref{RUT}) has two invariants: $\left\vert \mathbf{r}\left(
t\right) \right\vert =const,\ \mathbf{u}\left( t\right) \cdot \tfrac{d\, 
\mathbf{r}\left( t\right) }{dt}=0$ which follow from the structure of (\ref%
{rt}).

\section{Computational results}

We have performed several computations for the Bloch equation (\ref{RUT})
and the discrete-time dynamical system (\ref{QRQsimpl}), parameterized as
described in Section 3.2, to show dynamics of the Bloch equation and to
demonstrate how the map (\ref{QRQsimpl}) samples the flow of Eq.(\ref{RUT}).
Exact solutions of the map (\ref{QRQsimpl}) as well as of the Bloch equation
(\ref{RUT}) are given by (\ref{EXACTa}) and (\ref{EXACTb}), respectively.

Since $\omega $ determines time scale only we put $\omega =1$. In all
computations described below vectors $\mathbf{p}$, $\mathbf{q}$ are
orthogonal, $\mathbf{p}=\left[ 1,\ 0,\ 0\right] $, $\mathbf{q}=\left[ 0,\
0,\ 1\right] $ and the initial condition is $\mathbf{r}_{0}=\left[ 0.6,\ 0,\
0.8\right] $. Motion on the sphere is bounded by parallels $A_{1,2}=\mp \max
\left\vert \mathbf{r}\left( \theta \right) \cdot \mathbf{q}\right\vert $
given by (\ref{BOUNDS}). For the present choice of $\mathbf{p}$, $\mathbf{q}$
, $\mathbf{r}_{0}$ we have $A_{1,2}=\mp 0.8$. In all figures below the
vector $\mathbf{q}$, parallels $A_{1,2}$ and the equator are plotted, where
thin dashed lines indicate points invisible for the observer.

The solution (\ref{EXACTa}), of discrete-time dynamical system (\ref%
{QRQsimpl}) with $Q$, $P$ given by (\ref{QP}) has been plotted in Fig.\ 1
for $\alpha =0.01$, $\beta =0.06$ $\left( \lambda =\beta /\left( 2\alpha
\right) =3\right) $ where angles $\alpha $, $\beta $ have been given in
degrees. The value of $\alpha $ is so small that points $\mathbf{r}_{0},\ 
\mathbf{r}_{2},\ \mathbf{r}_{4},\ \ldots $ lie so close one to another that
a seemingly continuous curve, sampling the flow of the Bloch equation (\ref%
{RUT}) very densely, has been obtained. The dynamics has been also generated
directly from (\ref{QRQsimpl}) with $R_{1}=QPR_{0}^{-1}Q^{-1}$ (the angle $%
\chi _{0}\neq 0$ and arbitrary otherwise) to the same effect, the unit
vectors $\mathbf{r}_{N}$ have been renormalized after each iteration to
avoid numerical instabilities. The whole trajectory has three-fold symmetry
with respect to the $\mathbf{q}$ axis.

\begin{figure}[!ht]
\begin{equation*}
\includegraphics[width=6.4 cm]{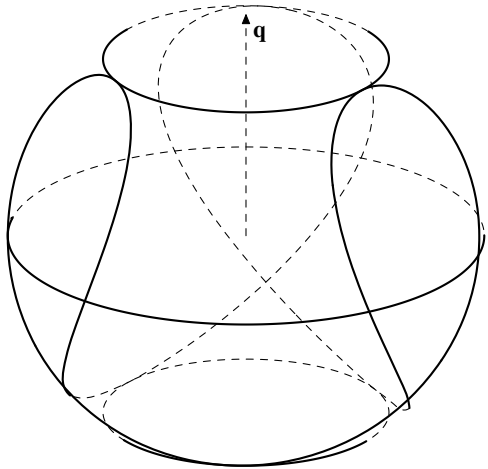}
\end{equation*}%
\caption{The Bloch equation (\protect\ref{RUT}) and discrete-time dynamical
system (\protect\ref{QRQsimpl}), $\protect\alpha =0.01$, $\protect\beta %
=0.06 $, $\protect\lambda =3$.}
\end{figure}

In Fig. 2 dynamics of vectors $\mathbf{r}_{N}$ obtained from (\ref{EXACTa})\
has been plotted for $\alpha =2$, $\beta =8$ ($\lambda =\beta /\left(
2\alpha \right) =2$). We thus obtain forty five points marked with dots. The solution (\ref{EXACTa}%
) for $\alpha =0.01$, $\beta =0.04$ ($\lambda =\beta /\left( 2\alpha \right)
=2$) sampling the Bloch equation densely has been also plotted. The closed
curve has two-fold symmetry with respect to the $\mathbf{q}$ axis.

\begin{figure}[!ht]
\begin{equation*}
\includegraphics[width=6.4 cm]{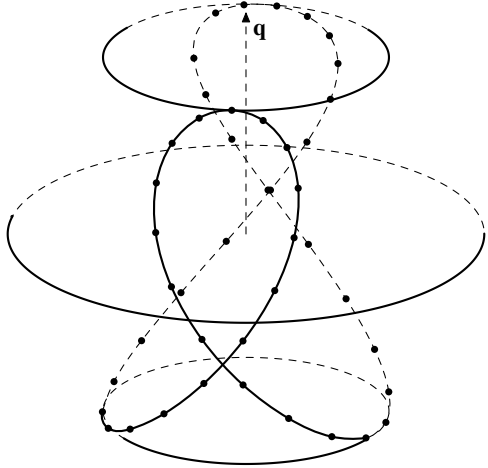}
\end{equation*}%
\caption{The Bloch equation (\protect\ref{RUT}) and discrete-time dynamical
system (\protect\ref{QRQsimpl}), $\protect\alpha =2$, $\protect\beta =8$
(dots) and $\protect\alpha =0.01$, $\protect\beta =0.04$ (solid and dashed
lines), $\protect\lambda =2$.}
\end{figure}

\begin{figure}[!ht]
\begin{equation*}
\includegraphics[width=6.4 cm]{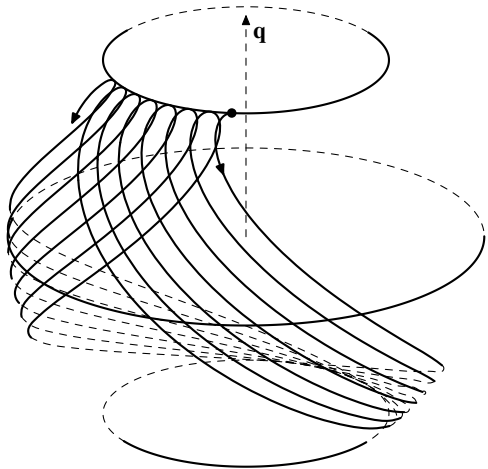}
\end{equation*}%
\caption{The Bloch equation (\protect\ref{RUT}) and discrete-time dynamical
system (\protect\ref{QRQsimpl}), $\protect\alpha =0.01$, $\protect\beta %
=0.0205$, $\protect\lambda =1.025$.}
\end{figure}

In Fig. 3 initial stage of dynamics of vectors $\mathbf{r}_{N}$ (\ref{EXACTa}
) has been plotted for $\alpha =0.01$, $\beta =0.0205$ ($\lambda =\beta
/\left( 2\alpha \right) =1.025$), dot on the upper parallel marks the
initial vector $\mathbf{r}_{0}$. Had the value of $\lambda $ be equal to one exactly the trajectory would
consist of one closed loop. Since $\lambda =1.025=41/40$ the whole closed
curve consists of forty one loops (in Fig. 3 seven such loops have been
shown) and samples densely the Bloch equation. In the case of close but
irrational value of $\lambda $, e.g. $\lambda =\sqrt{2}-0.389=1.0252\ldots $
, the trajectory is ergodic on the whole spherical sector bounded by two
parallels $A_{1,2}=\mp 0.8$.

\section{Summary and discussion}

We have introduced in Subsection 3 a class of discrete-time invertible maps (%
\ref{QRQ}) on an arbitrary group $\mathcal{G}$ and the exact solution (\ref%
{SOL}) of this map has been found. Maps of form (\ref{QRQ}), parameterized
on a Lie group, generate points in the parameter space which sample a
trajectory in this space. This curve can be generated forward as well as
backward from a given initial condition. This suggests that the group action
(\ref{QRQ}) may correspond to a flow of a differential equation. The results
described above generalize significantly our earlier findings \cite%
{Okninski1996b}.

In Section 3.2 a simplified dynamics (\ref{QRQsimpl}) has been parameterized
on the unit sphere, i.e. for $\mathcal{G}=SU\left( 2\right) $, and it has
been shown directly by constructing the solution (\ref{2k}), (\ref{EXACTa})
and (\ref{EXACTb})\ that the maps samples the flow of the Bloch equation (%
\ref{RUT}) exactly and arbitrarily densely. It can be thus stated that the
Bloch equation (\ref{RUT}) is a unique suspension of the map (\ref{QRQsimpl}%
).

In the special case $\mathbf{p}\cdot \mathbf{q=0}$ we recover the solution
obtained by H.K. Kim and S.P. Kim \cite{Kim2006}, see also Kobayashi papers
where several methods to solve the Bloch equation were described earlier 
\cite{Kobayashi2002, Kobayashi2003, Kobayashi2004}. It seems that the
present formulation leads to some progress in understanding the Bloch
equation. The role of the parameter $\lambda $ has been elucidated in
Subsection 3.2. More exactly, it follows from the condition (\ref{periodic})
that for fixed rational value of parameter $\lambda =\beta /\left( 2\alpha
\right) $ the flow of Eq.(\ref{RUT}) generates closed trajectories
consisting of a finite set of points. For \emph{\ }$\alpha ,\beta
\rightarrow 0$ and fixed value of $\lambda $ points obtained according to (%
\ref{EXACTa}) sample trajectories of (\ref{RUT}) exactly and arbitrarily
densely. Moreover, it has been shown in Section 3.3 that dynamical system (%
\ref{QRQsimpl}) has rotational symmetry with respect to the $\mathbf{q}$\
axis from which dynamical restrictions for the quantity $\mathbf{r}\left(
\theta \right) \cdot \mathbf{q}$ have been derived, see (\ref{BOUNDS}). It
follows from (\ref{EXACTb}) that for $\lambda =m/n$, with $m$, $n$
relatively prime, the curve $\mathbf{r}\left(\theta \right) $ has $m$ - fold
symmetry with respect to the $\mathbf{q}$\ axis. All these results have been
visualized in Section 5 where computational results have been presented. The
figures have been produced by code written in the MetaPost picture-drawing
language \cite{Hobby1995}.

The results described in the present paper can be generalized in several
directions. First of all it should be determined when the general group
dynamical system (\ref{QRQ}) samples a continuous curve. Whenever this is
the case it should be possible to construct from a solution (\ref{SOL}),
computed for some $Q_{N}$'s, an ODE for which the group action is a
stroboscopic map. Of course, uniqueness of such construction should be
investigated. A very simple choice of $Q_{N}$'s is $Q_{2K}=S$, $Q_{2K+1}=T$, 
$K=0,1,\ldots $ . We found in our early computations that for $ST\neq TS$
dynamics of Eq.(\ref{QRQ}) was very complicated \cite{Okninski1993}, yet
there is a closed form solution (\ref{SOL}). Finally, group actions on other
groups such as $SL_{2}\left( R\right) $ can be considered.

\section*{Acknowledgement}

It is a pleasure to thank Andrzej Lenarcik for introducing the author to the
graphics programming language MetaPost.

\end{subequations}


\begin{thebibliography}{99}
\bibitem{Okninski1992} A. Okninski, Physica D \textbf{55}, 358 (1992).

\bibitem{Okninski1994} A. Okninski, Int. J. Bifurcation and Chaos \textbf{4}
, 209 (1994).

\bibitem{Ahmed1994} E. Ahmed, A.E.M. El-Misiery, Int. J. Theor. Phys. 
\textbf{33}, 1681 (1994).

\bibitem{Okninski1996a} A. Okninski, A. Kowalska, J. Tech. Phys. \textbf{37}
, 395 (1996).

\bibitem{Shimizu1963} H. Shimizu, Ann. of Math. \textbf{77}, 33 (1963).

\bibitem{Leutbecher1967} A. Leutbecher, Math. Zeit. \textbf{100}, 183 (1967).

\bibitem{Brooks1981} R. Brooks, J.P. Matelski, in \textit{Riemann Surfaces
and Related Topics: Proceedings of the 1978 Stony Brook Conference}, edited
by I.Kra, B. Maskit, \textit{Ann. Math. Studies} \textbf{97} (Princeton
Univ. Press, Princeton, N.J., 1981), p. 65.

\bibitem{Gehring1989} F.W. Gehring, G.J. Martin, Bull. Am. Math. Soc. 
\textbf{21}, 57 (1989).

\bibitem{Mumford2002} D. Mumford, C. Series, D. Wright, \textit{Indra's
pearls: The vision of Felix Klein }(Cambridge University Press, Cambridge,
2002).

\bibitem{Kowalski1996} K. Kowalski, J. Rembielinski, Physica D \textbf{99},
237 (1996); chao-dyn/9801019.

\bibitem{Kowalski1998} K. Kowalski, J. Rembielinski, Chaos, Solitons and
Fractals \textbf{9}, 437 (1998); chao-dyn/9801020.

\bibitem{Aldrovandi1998} R. Aldrovandi, L.P. Freitas, J. Math. Phys. \textbf{%
\ 39}, 5324 (1998).

\bibitem{Gralewicz2000} P. Gralewicz, K. Kowalski, Chaos, Solitons and
Fractals \textbf{14}, 563 (2002); math-ph/0002044.

\bibitem{Tufillaro1992} N.B. Tufillaro, T. Abbott, J. Reilly, \textit{An
Experimental Approach to Nonlinear Dynamics and Chaos }(Addison-Wesley,
Redwood City, 1992).

\bibitem{deVries1993} J. de Vries, \textit{Elements of Topological Dynamics }
(Kluwer, Dordrecht, Boston, London, 1993).

\bibitem{Allen1975} L. Allen, J.M. Eberly, \textit{Optical Resonance and
Two-Level Atoms} (Wiley, New York, 1975).

\bibitem{Okninski1996b} A. Okninski, in \textit{Proceedings of the
Conference: Nonlinearity, Bifurcation, Chaos: the Doors to the Future, 16-18
September 1996, Dobieszkow, Poland.}, p. 196.

\bibitem{Barut1977} A.O. Barut, R. Raczka, \textit{Theory of Group
Representations and Applications} (PWN - Polish Scientific Publishers,
Warszawa, 1977).

\bibitem{Gajdek1995} M. Gajdek, A. Okninski, \textit{Z}. Naukowe
Politechniki Swietokrzyskiej, Mechanika \textbf{54}, 39 (1995).

\bibitem{Pontryagin1966} L.S. Pontryagin, \textit{Topological groups}
(Gordon and Breach, New York, 1966).

\bibitem{Misner1973} C.W. Misner, K.S. Thorne, J.A. Wheeler, \textit{\
Gravitation} (W.H. Freeman and Company, San Francisco, 1973).

\bibitem{Kim2006} H.K. Kim, S.P. Kim, J. Kor. Phys. Soc. \textbf{48}, 119
(2006).

\bibitem{Kobayashi2002} M. Kobayashi, J. Math. Phys. \textbf{43}, 4654
(2002).

\bibitem{Kobayashi2003} M. Kobayashi, J. Math. Phys. \textbf{44}, 2331
(2003).

\bibitem{Kobayashi2004} M. Kobayashi, J. Math. Phys. \textbf{45}, 475 (2004).

\bibitem{Hobby1995} John D. Hobby, \textit{A User's Manual for METAPOST} (AT
\& T Bell Laboratories, 1995).

\bibitem{Okninski1993} A. Okninski, R. Rynio, Dynamics Days, Poznan, 9-12 VI
1993.
\end{thebibliography}
\end{document}